\documentclass[a4paper,twocolumn,11pt,unpublished]{quantumarticle}
\pdfoutput=1
\usepackage[utf8]{inputenc}
\usepackage[english]{babel}
\usepackage[T1]{fontenc}
\usepackage{amsmath}
\usepackage{hyperref}

\usepackage{tikz}
\usepackage{lipsum}

\usepackage{braket}
\usepackage{amssymb}
\usepackage{qcircuit}
\usepackage{subfigure}
\usepackage[numbers]{natbib}
\usepackage{float}

\begin{document}
	
	\title{Variational data encoding and correlations in quantum-enhanced machine learning}
	
	\author{Ming-Hao Wang}
	\email{wangmh@hubu.edu.cn}
	\affiliation{School of Physics, Hubei University, Wuhan 430062, China}
	
	\author{Hua Lu}
	\email{lvhuahg@163.com}
	\affiliation{School of Science, Hubei University of Technology, Wuhan 430068, China}
	
	\maketitle
	
\begin{abstract}
	Leveraging the extraordinary phenomena of quantum superposition and quantum correlation, quantum computing offers unprecedented potential for addressing challenges beyond the reach of classical computers. This paper tackles two pivotal challenges in the realm of quantum computing: firstly, the development of an effective encoding protocol for translating classical data into quantum states, a critical step for any quantum computation. Different encoding strategies can significantly influence quantum computer performance. Secondly, we address the need to counteract the inevitable noise that can hinder quantum acceleration.
	Our primary contribution is the introduction of a novel variational data encoding method, grounded in quantum regression algorithm models. By adapting the learning concept from machine learning, we render data encoding a learnable process.
	Through numerical simulations of various regression tasks, we demonstrate the efficacy of our variational data encoding, particularly post-learning from instructional data. Moreover, we delve into the role of quantum correlation in enhancing task performance, especially in noisy environments. Our findings underscore the critical role of quantum correlation in not only bolstering performance but also in mitigating noise interference, thus advancing the frontier of quantum computing.
\end{abstract}
	
\section{Introduction}

	Quantum information processing stands at the forefront of next-generation information technology, offering the potential for exponential speedups over classical counterparts \cite{Nielsen2010}. Landmark algorithms like Shor's for large number factoring and the Harrow-Hassidim-Lloyd algorithm for linear systems exemplify this potential within the standard gate-based quantum computation model \cite{Shor1997, Harrow2009}. However, the current landscape is dominated by noisy-intermediate-scale quantum (NISQ) devices, characterized by their hundreds of noisy qubits and the consequent limitations in achieving large-scale, fault-tolerant quantum computing \cite{Preskill2018}. This reality steers contemporary research towards designing algorithms suitable for NISQ devices that still exploit quantum advantages \cite{Bharti2022,Callison2022}.

	In this context, hybrid quantum-classical algorithms (HQCAs) have emerged as a promising approach, demonstrating success in various applications ranging from calculating eigenstates of physical Hamiltonians to optimization and classification tasks \cite{Cerezo2021, Zhang2022, Jones2019, Liu2021a, Xu2021, Kandala2017, Chalumuri2021}. HQCAs, akin to machine learning algorithms, involve training computers to recognize patterns and minimize cost functions \cite{Jordan2015}. They utilize parameterized quantum circuits (PQCs) or quantum neural networks, widely adopted in quantum machine learning (QML) for tasks like classification and generative modeling \cite{Benedetti2019,Du2020,Nghiem2021,LaRose2020,Schuld2020,DallaireDemers2018,Lloyd2018,Gao2022}. The potential synergy between machine learning and quantum computing is a burgeoning area of interest \cite{Biamonte2017,Dunjko2016,Dunjko2018}.

	Central to HQCAs is the encoding of classical information into quantum states, a process known as quantum feature mapping (QFM) \cite{Havlicek2019,Schuld2019}. This encoding leverages the expressiveness of large Hilbert spaces to enhance data processing for quantum advantage \cite{Dunjko2016}. However, the choice of encoding schemes is critical, as it can significantly impact the performance of quantum algorithms. This paper delves into the concept of variational data encoding (VDE) within the realm of quantum regression algorithms. We explore the trainability of VDEs using PQCs, which are adjustable in quantum-classical optimization loops. Our investigation extends to the role of quantum correlation in training VDEs, suggesting its potential to enhance capacity and resist local noise.

\section{Variational data encoding}

	This section provides a comprehensive overview of QFMs, a cornerstone in quantum computation. QFMs are instrumental in encoding an input vector $x$ into a corresponding quantum state $\ket{\psi(x)}$. From a mathematical perspective, this encoding is represented as a mapping:
	\begin{equation}\label{eq:encoding}
		\psi:x\mapsto \ket{\psi(x)},
	\end{equation}
	where $x$ is an $N$-dimensional vector belonging to the real vector space $\mathbb{R}^N$, and $\ket{\psi(x)}$ denotes a quantum state within a $d$-dimensional Hilbert space $\mathcal{H}$.

	Physically, QFMs are implemented through a data-dependent quantum circuit. This circuit, characterized by a unitary operation $E(x)$, transforms a standard initial state $\ket{0}$ into the desired quantum state:
	\begin{equation}\label{eq:physical_encoding}
		\ket{\psi (x)} = E(x)\ket{0}.
	\end{equation}
	Typically, QFMs exhibit nonlinearity and the dimensionality of the quantum states' Hilbert space often surpasses that of the input space. This characteristic allows for a more expressive representation of data in quantum computing.

	QFMs are closely related to kernel methods in machine learning \cite{Hofmann2008, Lloyd2020, Jerbi2023}. The trick for both is to map the input vectors into a higher-dimensional feature space in which the feature vectors are easier to analyze. The common quantum encoding schemes, such as basis encoding and amplitude encoding, fall into this category. We refer interested readers to the reference and its supplemental material for details \cite{Schuld2019}.

	To make QFMs learnable, we further parameterize QFMs by introducing a parameter $\xi=(\xi_1,\xi_2,\cdots)$, denoting them as $E_{\xi}(x)$. We emphasize that $\xi$ is fixed for a specific QFM. Changing it results in a new QFM. By contrast, $x$ is the input vector to be encoded and it varies with different inputs.
	We call this quadratic parameterized quantum circuit variational data encoding. To find the right parameter $\xi$, a classical optimizer is needed. The optimization process consists of two parts. First, the quantum hardware is run and outputs measurements. Second, the classical optimizer calculates the loss function and its gradient based on these measurements, and updates the parameter $\xi$.
	After optimization, we hope to obtain a high-performance QFM such that the encoded states $\ket{\psi(x)}=E_{\xi}(x)\ket{0}$ is suitable for subsequent quantum computation to reveal quantum advantage. The optimization will be further detailed in Sec. \ref{opt}.

\section{Variational Quantum Regression Algorithms}

In the context of VDE, we introduce Variational Quantum Regression Algorithms (VQRAs) and provide a detailed exploration of their theoretical framework.
\subsection{Theory}
Consider a dataset, often referred to as training data in machine learning, consisting of $ M $ data points, denoted as $ \{x^{(m)}, y^{(m)}\}_{m=1}^M $. This dataset is drawn from a specific input set $ \mathcal{X} $, where each $ x^{(m)} = \{x^{m}_1, x^{m}_2, \cdots, x^{m}_N\}^T $ represents an $ N $-feature vector in the real vector space $ \mathbb{R}^N $, and $ y^{(m)} $ corresponds to a target value in $ \mathbb{R} $. The goal of a VQRA is to learn a functional mapping from $ x^{m} $ to $ y^{m} $ with as much accuracy as possible. This learned function is then utilized to predict the target value $ y $ for a new, previously unseen input point $ x $.

Assuming each $x^{(m)}$ has been transformed into a quantum state $ \ket{\psi(x^{(m)})} $ via a VDE circuit $ E_{\xi}(x^{(m)}) $, we have $ \ket{\psi(x^{(m)})} = E_{\xi}(x^{(m)})\ket{0} $. This transformation facilitates the definition of a complex kernel as follows:
\begin{equation}
	\kappa(x^{(m)},x^{(n)}) = \braket{\psi(x^{(m)})|\psi(x^{(n)})},
\end{equation}
which subsequently establishes a reproducing kernel Hilbert space $ R_\kappa $ \cite{Hofmann2008,Lloyd2020, Jerbi2023,Hubregtsen2022}. A function $ f(\cdot) $ within $ R_\kappa $ can be expressed through the inner products of the encoded state $ \ket{\psi(x)} $ and another state $ \ket{\Psi} $ in $ \mathcal{H} $, as:
\begin{equation}\label{rkhs}
	f(x) = \braket{\Psi |\psi(x)}.
\end{equation}
Selecting an appropriate VDE and state $ \ket{\Psi} $, $ f(\cdot) $ can be employed to approximate the desired input-output function. It is important to note that quantum state inner products are typically complex. However, a real kernel can be realized by either constraining the amplitudes of quantum feature states to real values or by considering the absolute value of the inner product \cite{Schuld2019}. In Section \ref{sec:simulation}, we adopt the latter strategy. Additionally, the magnitude of $ f(\cdot) $ is bounded between $ 0 $ and $ 1 $, necessitating suitable scaling when applicable.

To realize their full potential, VQRAs require training to identify the optimal parameters
$\xi$. This optimization process is conducted using a classical optimizer, which leverages a carefully crafted loss function. A common choice in regression tasks is the mean square error (MSE), defined as:
\begin{equation}\label{eq:loss_function}
	L=\frac{1}{M}\sum_{m=1}^M||f(x^{(m)})-y^{(m)}||^2+\Omega(f),
\end{equation}
where $\Omega(f)$ represents a regularization term aimed at reducing overfitting. This term is crucial in ensuring the generalizability of the model and maintaining a balance between fitting the training data and avoiding excessive complexity in the learned function.

In exploring the optimal function $\tilde{f}(x) =\braket{\Psi|\psi(x)}$, we assume the state $\ket{\Psi}$ can be decomposed as $\ket{\Psi} = \sum_{m=1}^{M} \alpha_m \ket{\psi(x^{(m)})}$. This allows us to express $\tilde{f}(x)$ as
\begin{equation}\label{eq:representer_1}
	\begin{split}
		\tilde{f}(x) =& \sum_{m=1}^{M}\alpha^*_m\braket{\psi(x^{(m)})|\psi(x)}\\
		=& \sum_{m=1}^{M}\alpha^*_m\kappa(x^{(m)},x).
	\end{split}
\end{equation}
If we define $\alpha^*_m = \beta_m y^{(m)}$ and substitute this into Eq. (\ref{eq:representer_1}), it yields:
\begin{equation}\label{eq:representer_2}
	\tilde{f}(x) = \sum_{m=1}^{M}\beta_m y^{(m)}\kappa(x^{(m)},x).
\end{equation}
We can interpret Eq. (\ref{eq:representer_2}) as follows:
\begin{enumerate}
	\item The state $\ket{\Psi}=\sum_{m=1}^{M}\beta_m y^{(m)}\ket{\psi(x^{(m)})} $ encapsulates all the training data through quantum superposition.
	\item The kernel function quantifies the similarity between pairs of data points in the input space.
	\item Predictions are made based on the weighted similarity between new data points and the existing training data, where the weights are given by $\beta_m$.
\end{enumerate}
Setting $\beta_m$ as a constant across all $m$ is a simplistic approach, implying equal weighting for all training data points. However, this method may overlook the nuances of the training data distribution and the inherent correlations among data points. Instead, optimizing these weightings could illuminate the underlying correlations, and this optimization can be effectively achieved using machine learning techniques.

\subsection{Implementation of Parameterized Quantum Circuits}
\begin{figure}[htbp]
	\centering
	\[
	\begin{array}{c}
		\Qcircuit @C=1.8em @R=1.5em{
			& \lstick{\ket{0}^{\otimes k}} & {/}\qw & \gate{M_\theta}  & \multigate{1}{Swap} & {/}\qw& \qw \\
			& \lstick{\ket{0}^{\otimes k}} & {/}\qw  & \gate{E_{\xi}(x)} & \ghost{Swap} & {/}\qw & \qw \\
			& \lstick{\ket{0}} & \qw  & \qw  & \ctrl{-1}  & \qw & \meter&  \\
		}	
	\end{array}
	\]
	\caption{Schematic representation of the quantum circuit utilized in VQRAs.
		It consists of three primary components: a memory circuit $M_\theta$, a VDE circuit $E_{\xi}(x)$,
		and a swap-test circuit. The memory and VDE circuits are applied to $k$ qubits initialized in state $\ket{0}^{\otimes k}$,
		while the swap-test circuit evaluates the overlap between the memory and encoded states.}
	\label{fig:framework}
\end{figure}
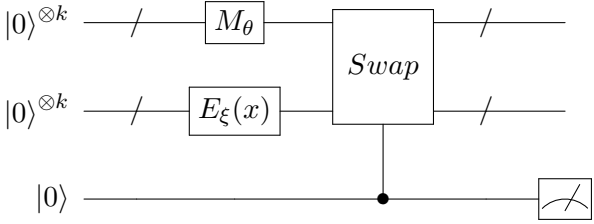

In this subsection, we delve into the implementation of PQCs for VQRAs, as illustrated in Fig. \ref{fig:framework}. The architecture comprises three main components: a memory circuit $M_\theta$, a VDE circuit $E_{\xi}(x)$, and a swap-test circuit \cite{Buhrman2001}. The memory circuit $M_\theta$, a PQC itself, is characterized by an adjustable parameter set $\theta=(\theta_1, \theta_2, \cdots)$. Its primary role is to learn and prepare the state $\ket{\Psi} = M_\theta\ket{0}$, encapsulating the dataset information.

The VDE circuit, another PQC, operates with two categories of parameters: trainable $\xi$ and the input vector $x$. The circuit encodes $x$ into a quantum state, where $\xi$ is fine-tuned during the optimization process. Both the memory and encoder circuits act on $k$ qubits initialized in the standard state $\ket{0}^{\otimes k}$.

The swap-test circuit's function is to evaluate the overlap between $\ket{\Psi}$ and $\ket{\psi(x^{(m)})}$, essentially computing the kernel function. According to quantum measurement theory, the probability of measuring `0' on the last qubit is given by $p(0) = \frac{1 + |\braket{\Psi|\psi(x^{(m)})}|^2}{2}$ \cite{Buhrman2001, GarciaEscartin2013, Zhao2021}, providing a direct way to quantify the similarity between quantum states.

\begin{figure*}[htbp]
	\centering
	\[
	\begin{array}{c}
		\Qcircuit @C=1.8em @R=1.5em{
			& \lstick{\ket{0}}& \gate{R_x(\theta_1^0)}& \ctrl{1} & \qw &\qw & \targ &\gate{R_x(\theta_1^d)} & \qw & \qw\\
			& \lstick{\ket{0}}& \gate{R_x(\theta_2^0)}& \targ    & \ctrl{1} & \qw&\qw & \gate{R_x(\theta_2^d)} & \qw & \qw\\
			& \lstick{\ket{0}}& \gate{R_x(\theta_3^0)}& \qw      & \targ    & \qw&\qw & \gate{R_x(\theta_3^d)} & \qw& \qw\\
			&&\vdots  & &&\vdots&&& \\
			& \lstick{\ket{0}}& \gate{R_x(\theta_k^0)}&\qw       &\qw  & \qw& \ctrl{-4}     &\gate{R_x(\theta_k^d)}  &\qw& \qw\\
			& \hspace{45em}_{\times D_M}\gategroup{1}{4}{5}{8}{2em}{(} \gategroup{1}{4}{5}{8}{2em}{)}
		}	
	\end{array}
	\]
	\caption{Schematic representation of the memory circuit $M_{\theta}$ utilized in VQRAs.
		The circuit consists of $D_M$ layers, each comprising a sequence of single-qubit rotations $R_{x}(\theta)$ along the X-axis and controlled-NOT gates.
		Here, $R_{x}(\theta_i^j)$ represents the rotation operator for the $i$-th qubit in the $j$-th layer, and the combination of these rotations and entanglements
		forms the full structure of the circuit. This architecture is designed to enhance the expressivity and capacity of the memory circuit for encoding quantum information.}
	%	\caption{Architecture for the memory circuit $M_{\theta}$. $R_{x}(\theta)$ denotes a single-qubit rotation operator along the X-axis with angle $\theta$. }
	\label{fig:m_circuit}
\end{figure*}
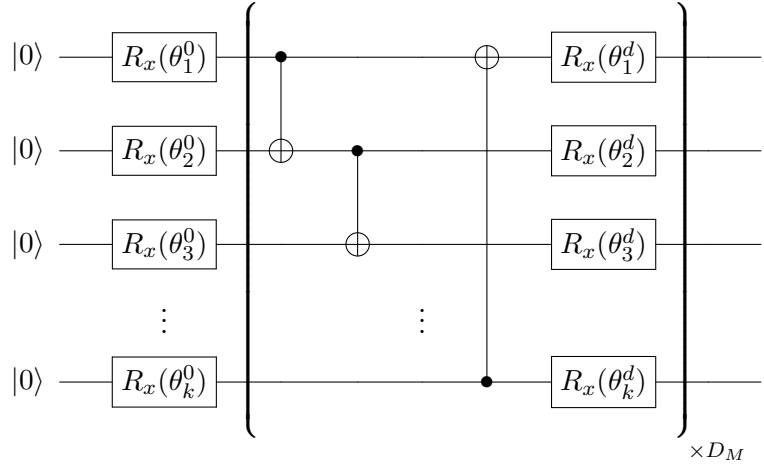

The construction of a PQC can vary significantly based on the specific research focus or the constraints of the quantum hardware \cite{Tang2021, Kandala2017}. For the memory circuit $M_\theta$, we design it using $D_M$ layers of a fundamental unit, each sharing the same structural framework. Each of these units is composed of single-qubit rotations $R_x(\theta_i)$ and controlled-NOT gates, as depicted in Fig. \ref{fig:m_circuit}. Here, $D_M$ represents the depth of $M_\theta$, and increasing $D_M$ correlates with enhanced expressivity of the memory circuit, allowing for more complex quantum state representations \cite{Du2020,Schuld2021}.

\begin{figure*}[htbp]
	\[
	\begin{array}{c}
		\Qcircuit @C=1.8em @R=1.5em{
			& \lstick{\ket{0}}
			&\gate{R_y(\xi_1^{d,1})}&\multigate{4}{E^{d,1}}&\gate{R_{x}( x_{1})}&\qw&\cdots&&\gate{R_y(\xi_1^{d,N})}&\multigate{4}{E^{d,N}}&\gate{R_{x}( x_{N})} &\qw& \qw\\
			& \lstick{\ket{0}}
			&\gate{R_y(\xi_2^{d,1})}& \ghost{E^{d,1}}      &\gate{R_{x}( x_{1})}&\qw&\cdots&&\gate{R_y(\xi_2^{d,N})}&\ghost{E^{d,N}}       &\gate{R_{x}( x_{N})}
			&\qw& \qw\\
			&\lstick{\ket{0}}
			&\gate{R_y(\xi_3^{d,1})}& \ghost{E^{d,1}}      &\gate{R_{x}( x_{1})}&\qw&\cdots&&\gate{R_y(\xi_3^{d,N})}&\ghost{E^{d,N}}       &\gate{R_{x}( x_{N})}
			& \qw& \qw\\
			& &\vdots&&\vdots &&\cdots&&\vdots&&\vdots\\
			&\lstick{\ket{0}}
			&\gate{R_y(\xi_k^{d,1})}& \ghost{E^{d,1}}      &\gate{R_{x}( x_{1})}&\qw&\cdots&&\gate{R_y(\xi_k^{d,N})}&\ghost{E^{d,N}}       &\gate{R_{x}( x_{N})}
			&\qw& \qw\\
			& \hspace{80em}_{\times D_E}\gategroup{1}{3}{5}{11}{1em}{(} \gategroup{1}{3}{5}{11}{1em}{)}
		}
	\end{array}
	\]
	\caption{Schematic representation of the VDE circuit $E_{\xi}(x)$. The circuit is composed of $D_E$ layers,
		each containing a series of rotation gates $R_{y}(\xi_i^{d,j})$ and entanglement operations $E^{d,j}$, followed by $R_{x}(x_j)$ rotations.
		Here, $\xi$ represents the set of trainable parameters defining the encoding scheme, while $x$ denotes the input vector being encoded into the quantum state.
		This layered architecture allows for the flexible and expressive encoding of classical data into quantum states, catering to different encoding requirements and enhancing the circuit's overall capability.}
	\label{fig:e_circuit}
\end{figure*}
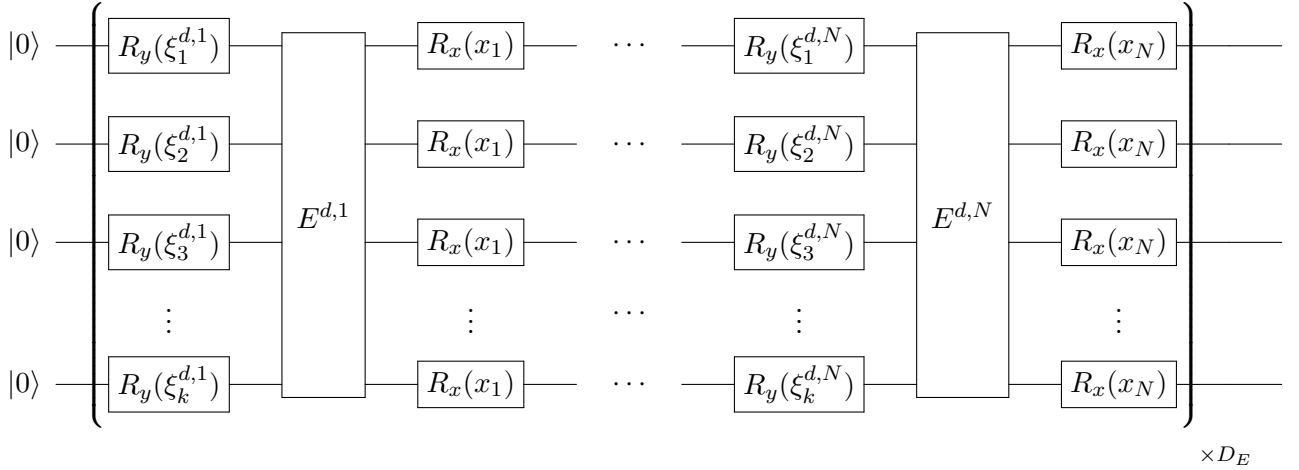

The construction of the VDE circuit follows a methodology similar to that of the memory circuit. Its detailed architecture is illustrated in Fig. \ref{fig:e_circuit}, with $D_E$ denoting the depth of the circuit $E_{\xi}(x)$. To foster entanglement within the circuit, we have designed the entanglement operations $E^{d,n}$ as follows:
\begin{widetext}
\begin{equation}
	E^{d,n}=e^{-iX_k X_1\xi^{d,n}_{2k}}e^{-iX_{k-1} X_k\xi^{d,n}_{2k-1}}\cdots e^{-iX_2 X_3\xi^{d,n}_{k+2}}e^{-iX_1 X_2\xi^{d,n}_{k+1}}, n=1,2,\cdots,N,
\end{equation}
\end{widetext}
where $X_i$ represents the Pauli-X operator acting on the $i$th qubit. The depth $D_E$ of the circuit $E\xi (x)$ plays a critical role in determining its expressivity. Notably, setting $D_E=1$ results in a traditional encoding scheme. In contrast, a depth of $D_E\ge 2$ corresponds to a data-reloading scheme, which has been shown to further enhance the expressivity of the circuit \cite{PerezSalinas2020}.

\subsection{Optimization}\label{opt}

Training is a crucial step for VQRAs to effectively make predictions on new data. Initially, the parameters of VQRAs are set to random values. During each iteration of the training loop, the quantum device is tasked with state preparation, processing, and finally, performing measurements on designated qubits. These repeated operations yield estimates of $f(x^{m})$. To determine the gradient of $f(x^{m})$, it may be necessary to run variations of the circuit.
These results, combined with a well-constructed loss function, enable a classical optimizer to refine the parameters within the quantum device. In our study, we employ MSE as the loss function, as detailed in Eq. (\ref{eq:loss_function}). This choice of loss function aligns with the objective of minimizing the deviation between the predicted and actual values, thus enhancing the accuracy and reliability of the VQRA's predictions.

QML encompasses a variety of learning rules, which can be broadly classified into gradient-free and gradient-based methods \cite{Mitarai2018, Schuld2019a}. The choice between these methods depends largely on the specific problem at hand, with each offering its own set of advantages and drawbacks. Gradient-based methods are often favored due to their rapid convergence and high precision, especially in scenarios involving a vast parameter space.
To acquire gradient information, several techniques have been developed \cite{Schuld2019a}. In our study, we opt for numerical differentiation as a straightforward approach to approximate gradients. This simplification allows us to utilize gradient-based optimization algorithms, such as Adam \cite{Kingma2014}, for updating the parameters in the PQCs. Employing such methods facilitates efficient and precise tuning of the circuit parameters, thereby enhancing the overall performance of the quantum machine learning model.

\section{Numerical simulations}\label{sec:simulation}

This section details the numerical simulations we conducted for VQRAs using the PennyLane software library. PennyLane, renowned for its versatility and open-source nature, serves as a comprehensive platform for quantum computing, quantum machine learning, and quantum chemistry. Its universal compatibility with various gate-based quantum computing platforms and simulators as backends makes it an ideal choice for a wide range of quantum algorithm implementations \cite{Bergholm2018}. This flexibility is particularly valuable, not only for educational and research demonstrations but also for the future application of these algorithms to practical problems \cite{Antipov2022}.

In our simulations, we focused on assessing the performance of VQRAs, exploring the quantum correlations present in the quantum states, and examining the impacts of noise on the system. These simulations provide insights into the efficacy and robustness of VQRAs under various conditions, offering valuable data for their further development and application.

\subsection{Performance}
In evaluating the performance of VQRAs, we initially focus on their ability to fit various functions. The quantum circuits are configured with $k=D_M=3$ and $D_E=6$, balancing complexity and computational feasibility.

Our simulations involved fitting tasks for several distinct functions: $f_1(x) = x^2$, $f_2(x) = e^x/e$, $f_3(x) = \sin^2(\pi x)$, and $f_4(x) = 1/(1+e^{-10x})$, all within the domain $x \in [-1,1]$. The training data for each function were uniformly distributed across this range. To mimic realistic data conditions, we introduced small Gaussian noise with a standard deviation of $\sigma = 0.01$ to the training data.

\begin{figure*}[ht]
	\centering
	\subfigure[$x^2$]{\includegraphics[width=0.35\textwidth]{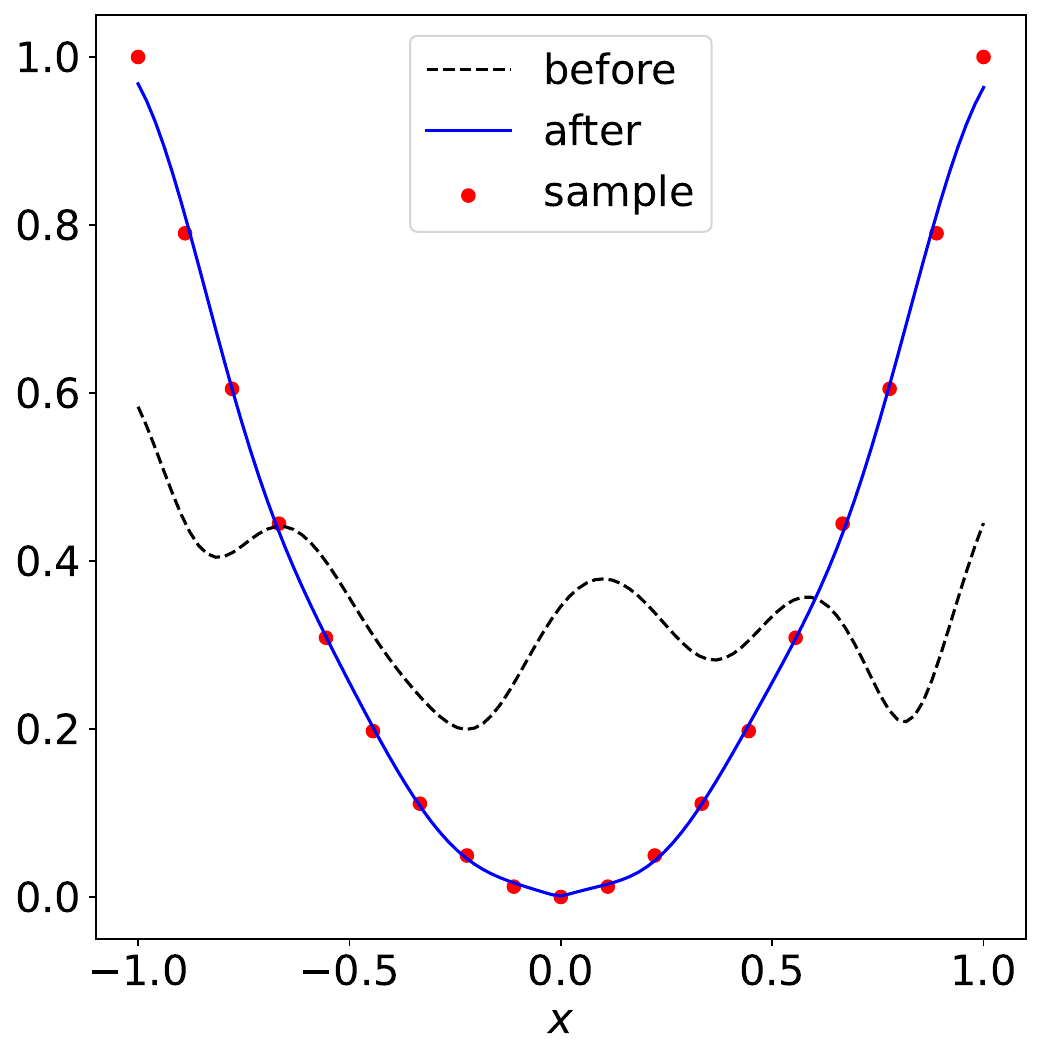}
		\label{fig:square}}
	\hfil
	\subfigure[$e^x/e$]{\includegraphics[width=0.35\textwidth]{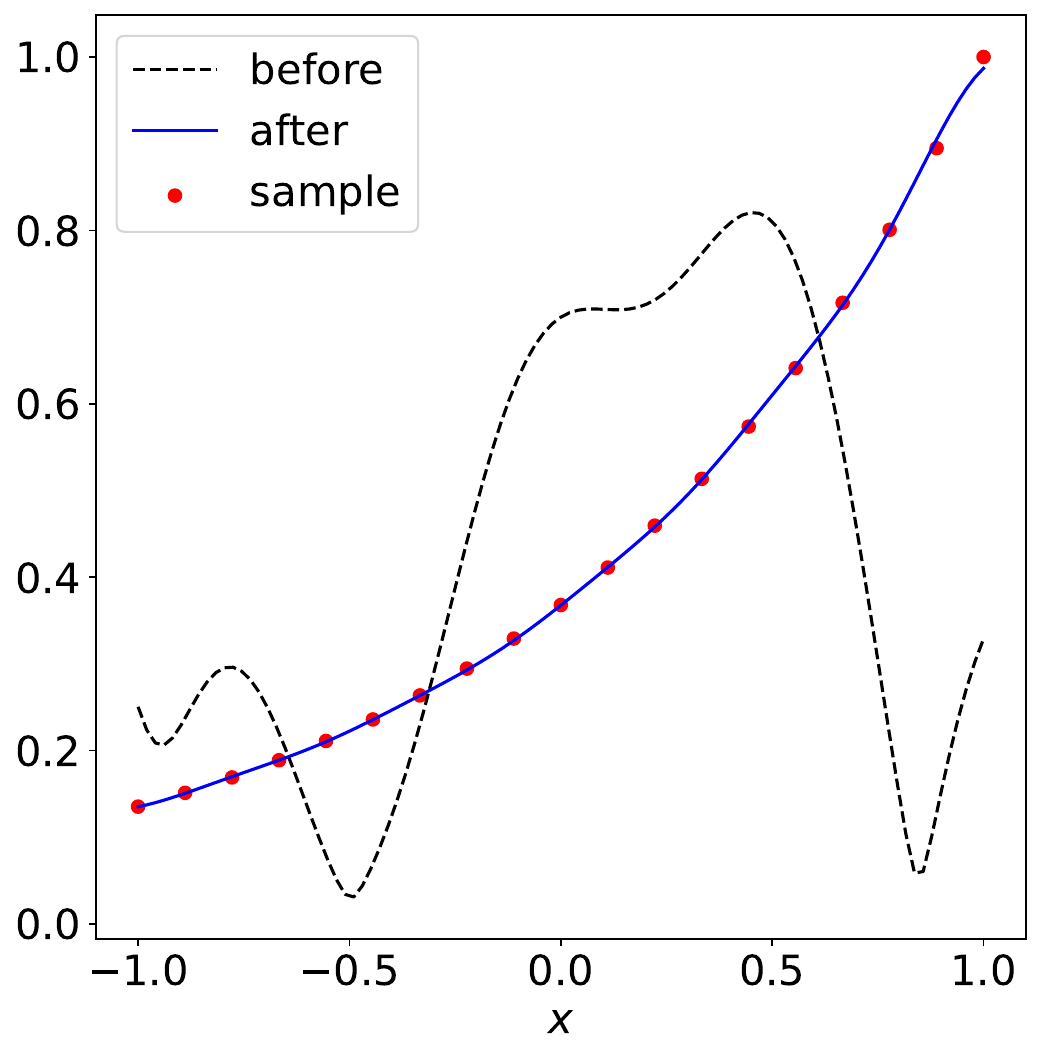}
		\label{fig:exp}}\\
	\subfigure[$\sin^2(\pi x)$]{\includegraphics[width=0.35\textwidth]{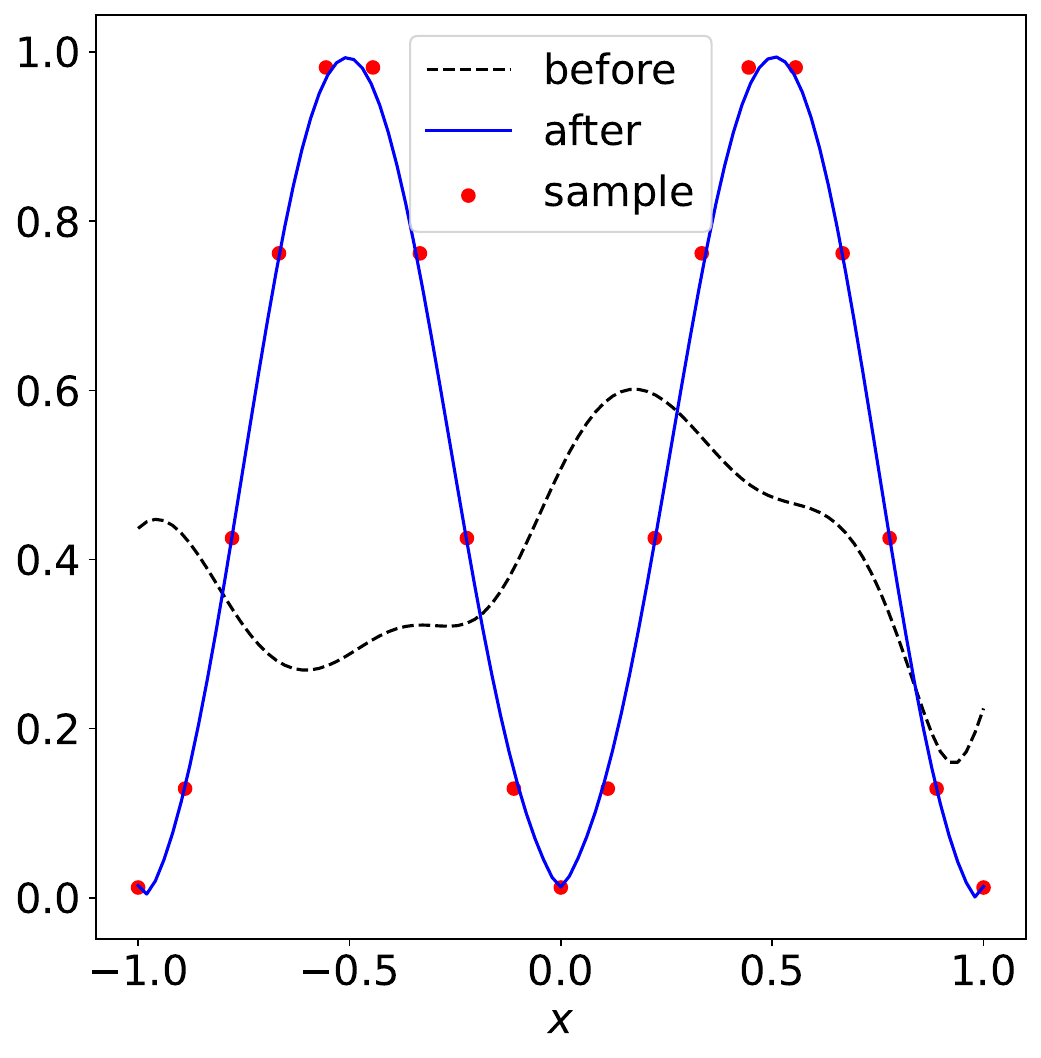}
		\label{fig:sin}}
	\hfil
	\subfigure[$1/(1+e^{-10x})$]{\includegraphics[width=0.35\textwidth]{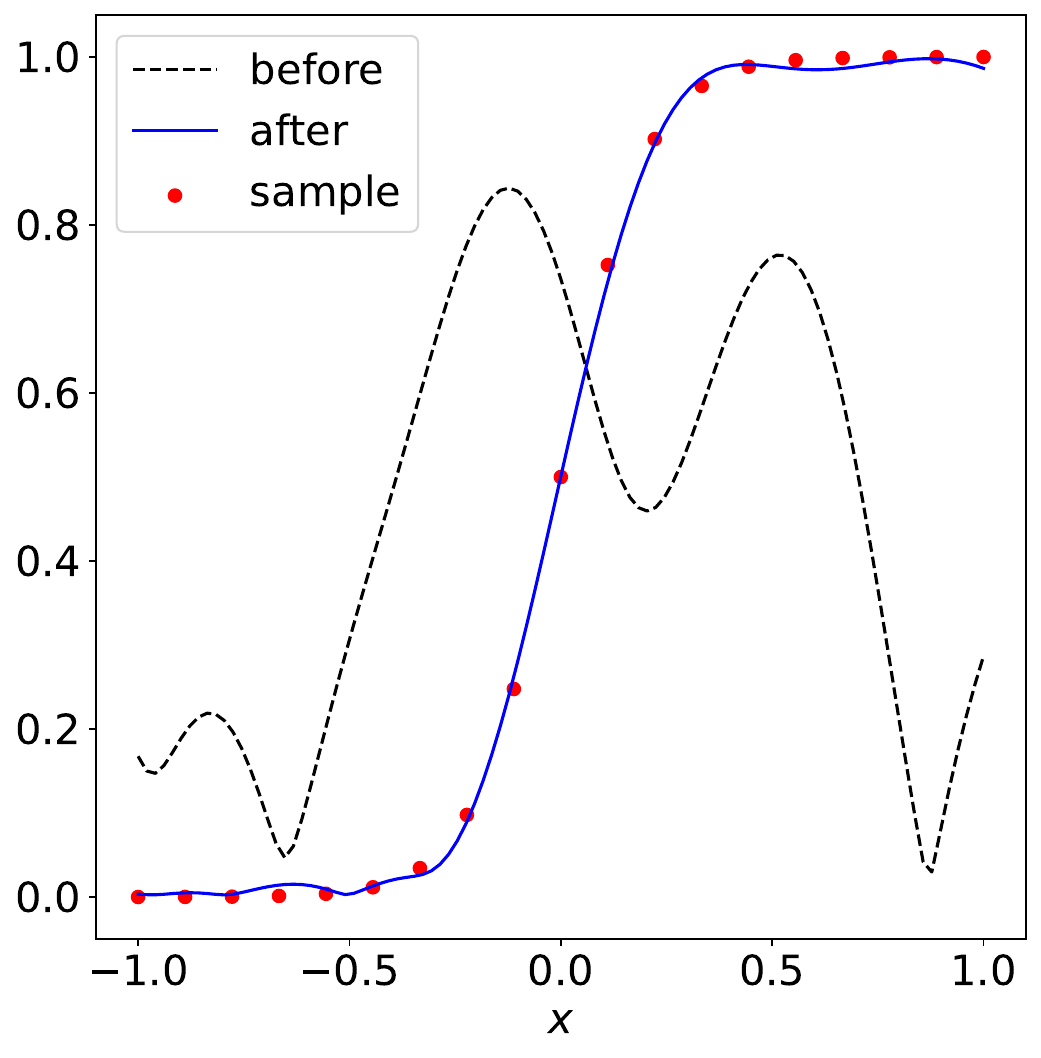}
		\label{fig:sigm}}
	\caption{Demonstration of VQRAs in fitting various single-variable functions.
		Panels (a) to (d) depict the fitting results for $x^2$, $e^x/e$, $\sin^2(\pi x)$, and $1/(1+e^{-10x})$ respectively.
		The quantum circuit parameters were configured with $k=D_{M}=3$ and $D_E=6$. Each function was approximated by training
		the VQRA model through 2000 iterations. These results highlight the algorithm's capability in accurately modeling different
		mathematical functions, showcasing its adaptability and precision.}
	\label{fig:one_var}
\end{figure*}

The outcomes of these simulations, illustrated in Fig. \ref{fig:one_var}, demonstrate that VQRAs, even with a limited number of qubits and shallow circuit depth, are capable of closely approximating all the tested functions. This success underscores the substantial expressive power inherent in the Hilbert space, even when harnessed by relatively simple quantum circuits.

\begin{figure*}[ht]
	\centering
	\subfigure[Samples]{\includegraphics[width=0.4\textwidth]{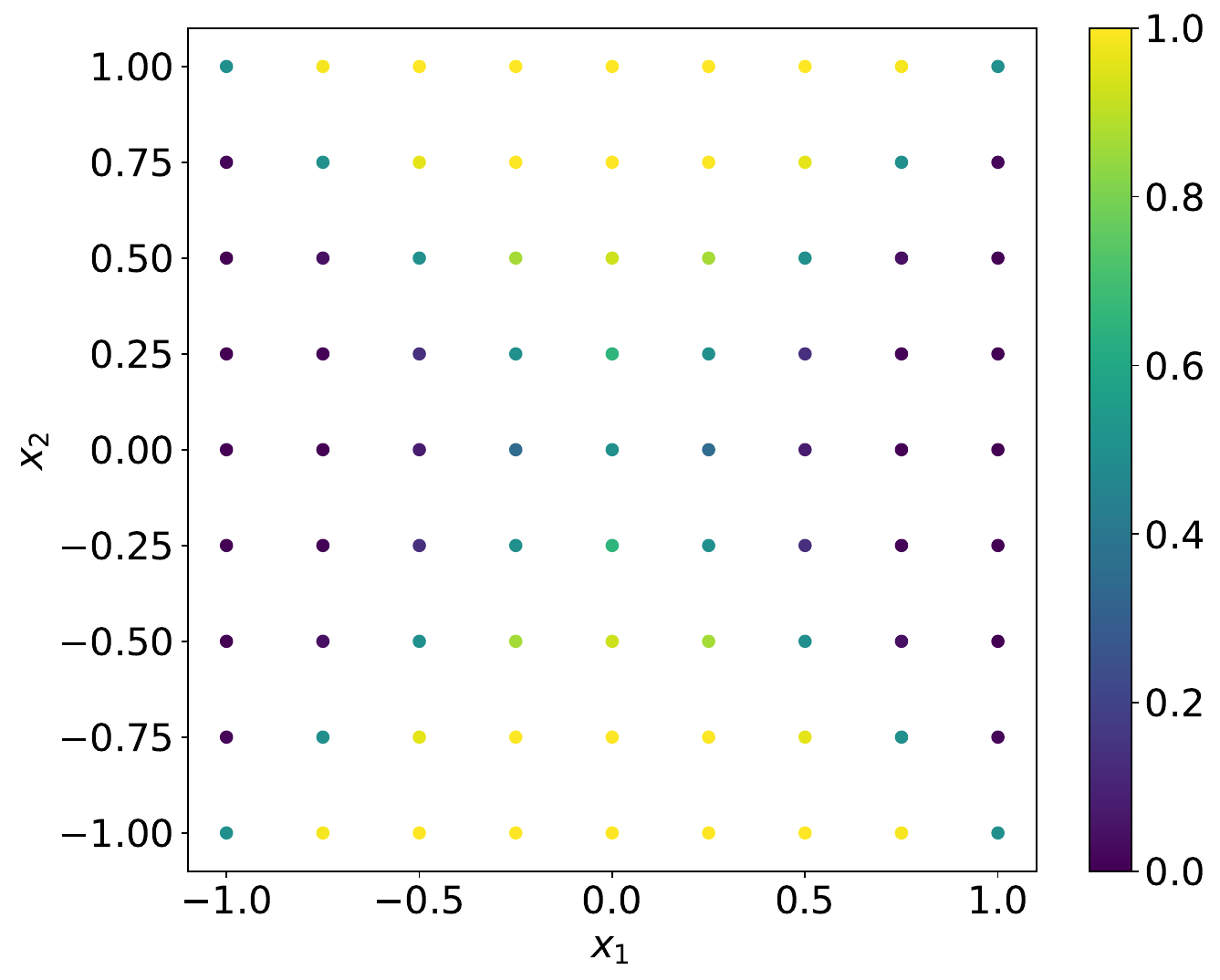}
		\label{fig:f5_sample}}
	\hfil
	\subfigure[Before]{\includegraphics[width=0.4\textwidth]{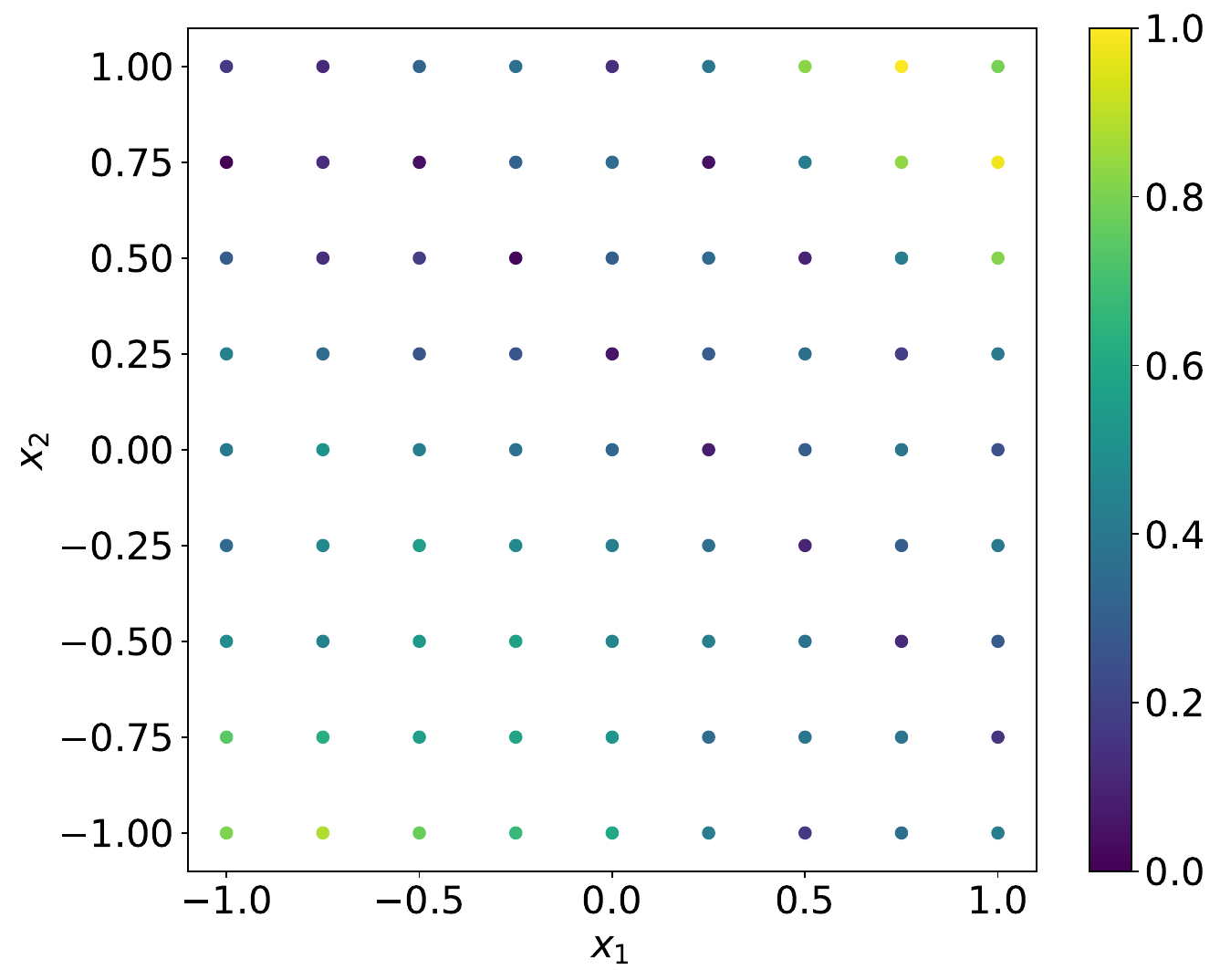}
		\label{fig:f5_before}}
	\\
	\subfigure[After]{\includegraphics[width=0.4\textwidth]{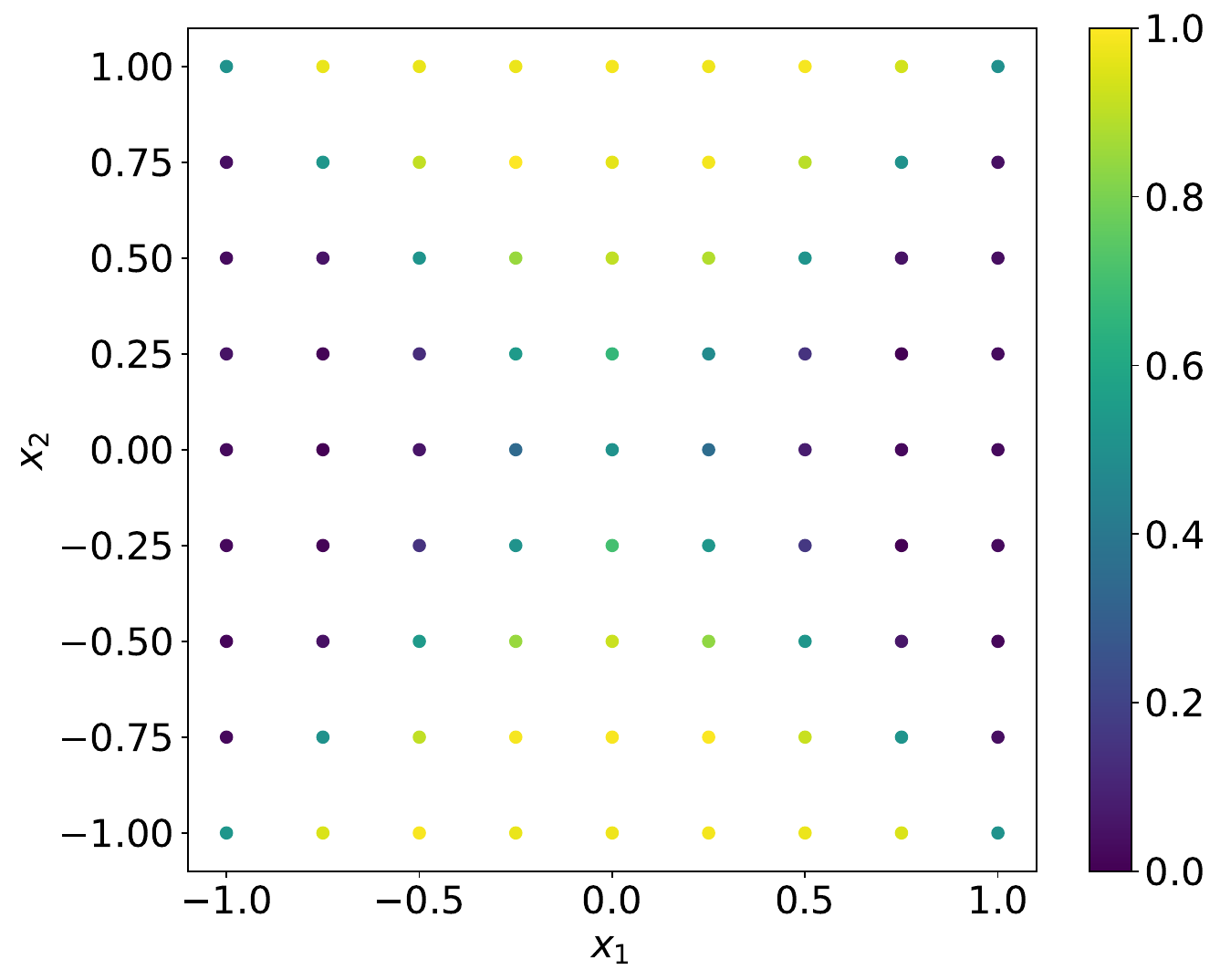}
		\label{fig:f5_after}}
	\hfil
	\subfigure[Loss]{\includegraphics[width=0.4\textwidth]{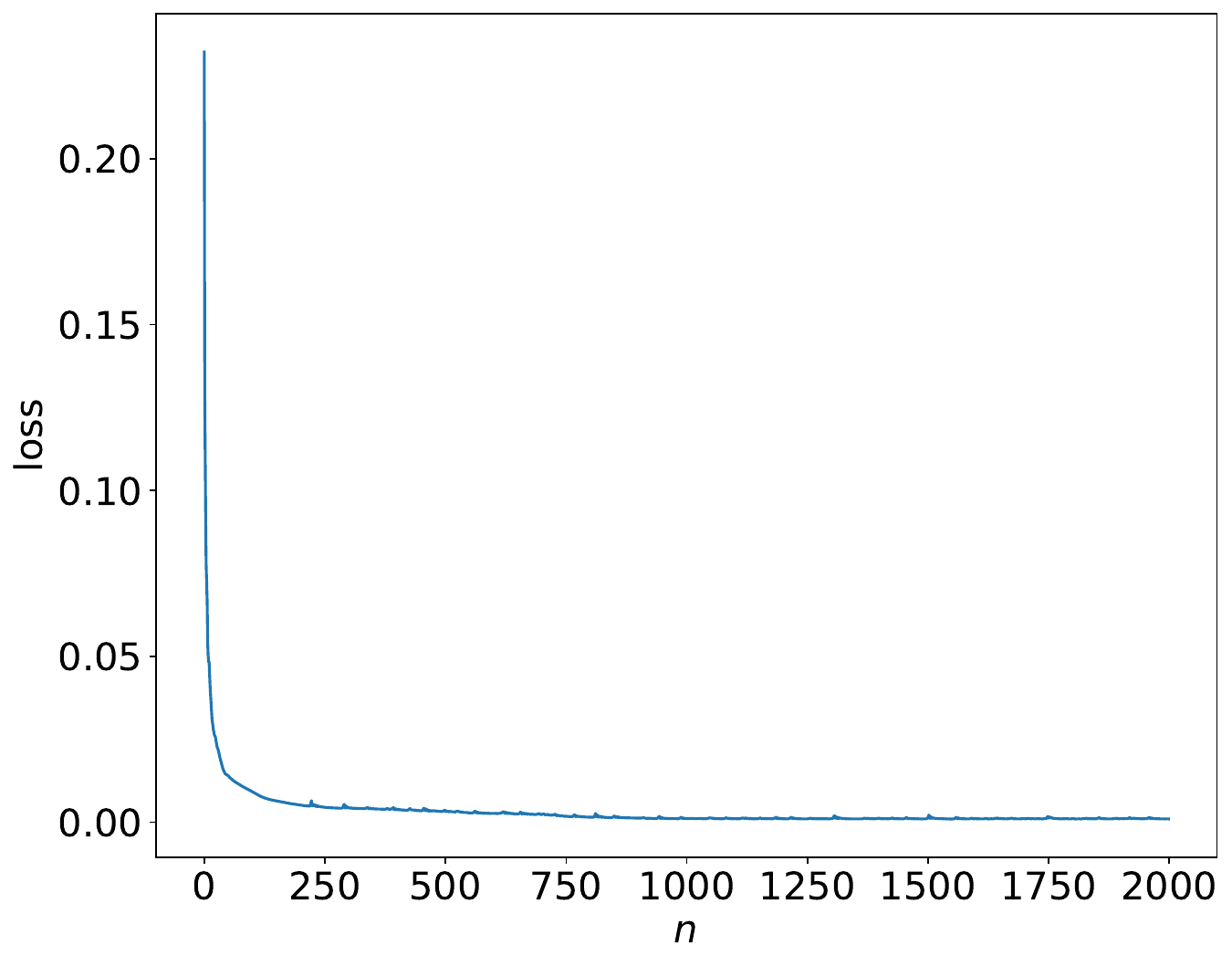}
		\label{fig:f5_loss}}
	\caption{Illustration of VQRA's efficacy in fitting the multivariate function $f_5(x_1, x_2) = \frac{1}{1+e^{10(x_1^2-x_2^2)}}$. The circuit parameters were set to $k=D_{M}=3$ and $D_E=6$. Panel (a) displays the training data samples. Panel (b) shows the initial model performance before training. Panel (c) demonstrates the improved fitting accuracy after 2000 training iterations. Panel (d) depicts the progression of the loss function throughout the training process. This sequence of images highlights the significant improvement in model performance post-training, showcasing the adaptability and learning capacity of VQRA for complex multivariate functions. }
	\label{fig:two_var}
\end{figure*}

In addition to single-variable functions, we extended our investigation to multivariate functions using VQRAs. An exemplary case is presented in Fig. \ref{fig:two_var}, where we demonstrate the fitting of the function $f_5(x_1,x_2) = \frac{1}{1+e^{10(x_1^2-x_2^2)}}$ within the domain $x_1,x_2 \in [-1,1]$. This result exemplifies the capability of VQRAs to accurately model functions involving multiple variables, highlighting their versatility and effectiveness in handling complex datasets.

Furthermore, the successful fitting of both $f_4$ and $f_5$ functions suggests that VQRAs can be adapted for quantum classification tasks. By setting appropriate classification boundaries, these algorithms can be trained to distinguish between different classes of data, as supported by existing studies in quantum classification \cite{Havlicek2019, Schuld2020}. This versatility makes VQRAs a promising tool not only for regression but also for classification problems in quantum machine learning.

\subsection{Quantum correlations in VQRAs}

Quantum correlations are a cornerstone of quantum information processing, with their unique properties offering significant advantages over classical systems \cite{Horodecki2009, Vedral2014, Erhard2020, Koehnke2021}. A deep understanding of these correlations, encompassing both their characterization and quantification, is essential for harnessing the full potential of quantum physics in practical applications.
In the context of VQRAs, quantum correlations manifest within both the memory and encoded states. While a universal metric for quantifying quantum correlations
in multi-body quantum systems remains elusive \cite{Modi2012, Adesso2016}, our study takes a qualitative approach to examine the influence of quantum correlations on VQRAs. We aim to elucidate how these correlations contribute to the performance and efficacy of the algorithm, thereby providing insights into the intricate interplay between quantum mechanics and machine learning.

\begin{figure}[ht]
	\centering
	\includegraphics[width=0.5\textwidth]{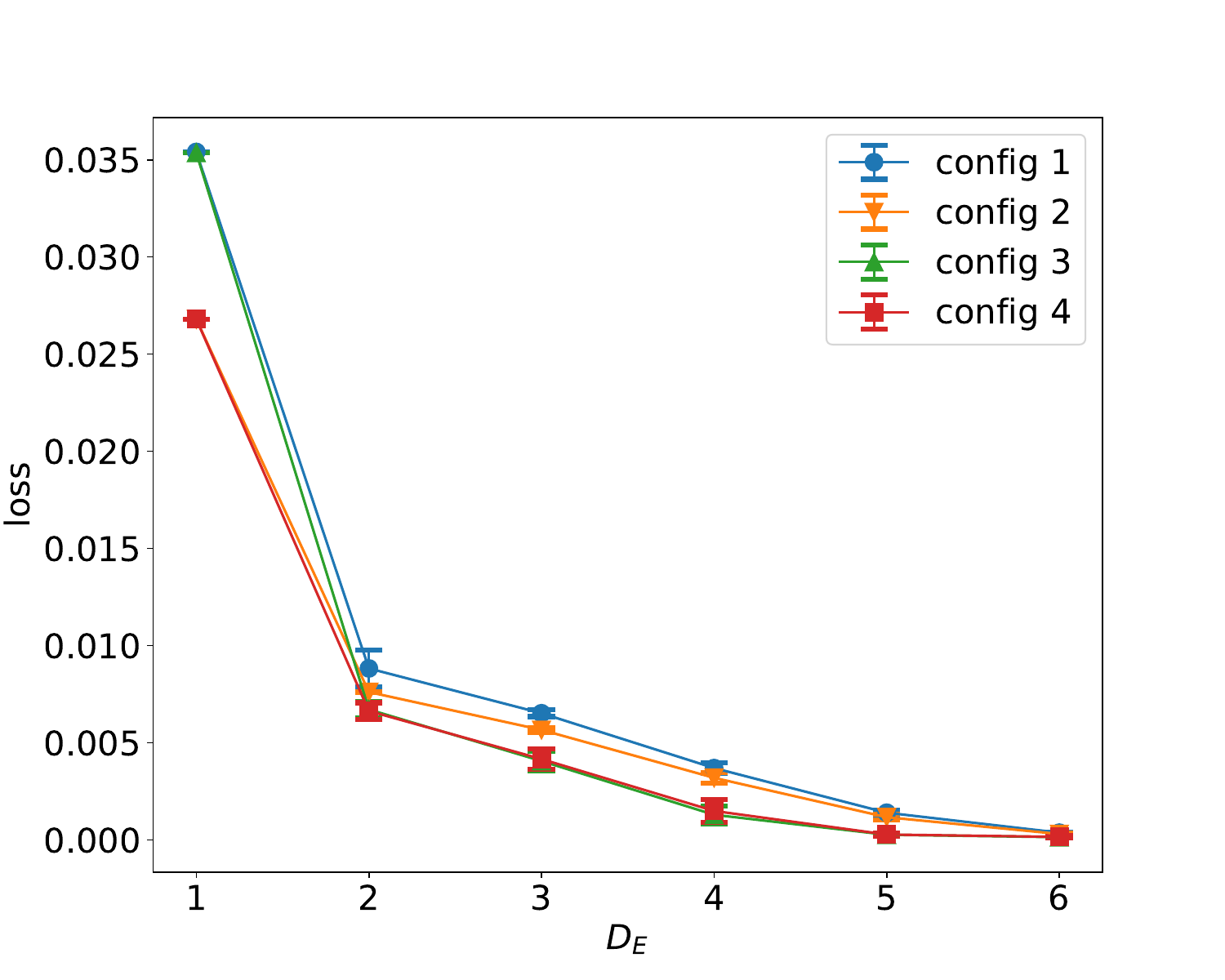}
	\caption{Comparative analysis of loss curves across different configurations and encoding depths in VQRAs. Here, the quantum circuit's structural parameters were configured with $k=D_M=3$. To investigate the influence of circuit depth on performance, we varied $D_E$ from 1 to 6 for each configuration. This figure highlights how different levels of quantum entanglement, influenced by the depth of the circuits, affect the loss and, consequently, the efficacy of the VQRA model. The results provide insights into the optimal balance between circuit depth and quantum correlation for efficient quantum information processing.}
	\label{fig:loss_config}	
\end{figure}

Initially, we examined VQRAs in an idealized, noise-free environment, focusing specifically on the fitting of function $f_4$. To delve deeper into the role of quantum correlations, we analyzed four distinct configurations of quantum circuits, each allowing for different levels of correlation. The configurations were as follows:
\begin{itemize}
	\item \textbf{Configuration 1:} All entanglement gates were removed from both the memory and the VDE circuits.
	\item \textbf{Configuration 2:} Entanglement gates in the memory circuit were retained, while those in the VDE circuit were removed.
	\item \textbf{Configuration 3:} Entanglement gates in the memory circuit were removed, but those in the VDE circuit were kept.
	\item \textbf{Configuration 4:} Both the memory and the VDE circuits retained their entanglement gates.
\end{itemize}
For each configuration, the depth $D_E$ was varied from 1 to 6. We conducted 10 training rounds for each circuit, with 2000 iterations per round. Post-training, we computed the average values and standard deviations of the loss function.
The results, depicted in Fig. \ref{fig:loss_config}, indicate that for shallow circuits, where quantum correlation is limited,
Configuration 4 consistently outperforms the other configurations. However, as the circuit depth increases, this advantage diminishes.
Deep circuits without correlations already exhibit satisfactory performance, obscuring the benefits of quantum correlations.
The impact of quantum correlations is more pronounced in medium-depth circuits, where increased leveraging of correlations corresponds with improved fitting accuracy.
For instance, in our simulations, we observed that $l_{1} > l_{2} > l_{3} \approx l_{4}$ when $D_E$ was set to 2, 3, 4, and 5, where $l_{i}$ represents the loss value.

In practical scenarios, encoding processes in VQRAs cannot be completely isolated from environmental interactions, leading to inevitable errors caused by noise. Such noise can be mathematically represented by noisy quantum channels with Kraus operators, defined as:
\begin{equation*}
	\varepsilon(\rho) = \sum_k E_k \rho E^{\dagger}_k,
\end{equation*}
where each $E_k$ is a Kraus operator satisfying $\sum_k E^{\dagger}_k E_k = I$. Given the myriad types of noisy quantum channels, we assume local and identical noise across all qubits for simplicity. This assumption is reasonable in scenarios where qubits are physically well-separated and uniformly affected by similar environmental factors.

We particularly focus on three prevalent types of noisy quantum channels for a single qubit: amplitude damping, phase damping, and symmetric depolarizing channels. These channels are characteristic of non-Markovian processes, leading to continuous loss of quantum information to the environment. Among them, symmetric depolarizing channels are noteworthy for their severe impact, replacing the qubit state with a completely mixed state at a probability $p$ \cite{Fontana2021}. The channel is mathematically modeled as:
\begin{equation}
	\begin{split}
		\varepsilon(\rho) =& (1-\frac{3p}{4})\rho + \frac{p}{4}(X\rho X+Y\rho Y+Z\rho Z)\\
		=& \frac{pI}{2}+(1-p)\rho,
	\end{split}
\end{equation}
where $X, Y, Z$ denote Pauli operators, and $p \in [0,1]$ represents the noise strength.

To simplify our analysis, we introduced noise channels immediately following the VDEs, resulting in the noisy encoded states:
\begin{equation}
	\tilde{\rho}{x^{(m)}} = \left(\otimes{i=1}^{k}\varepsilon_i\right)(\rho_{x^{(m)}}),
\end{equation}
where $\rho_{x^{(m)}} = \ket{\psi(x^{(m)})}\bra{\psi(x^{(m)})}$ is the noise-free encoded state for the input vector $x^{(m)}$, and $\varepsilon_i$ represents the single-qubit noisy channel acting on the $i$th qubit.

We conducted a detailed examination of the effects of noise on the effectiveness of VQRAs using the configurations outlined earlier. The structural parameters for these simulations were set to $k=D_M=3$ and $D_E=6$. We incrementally increased the intensity of the noise in each scenario. For each noise level, the circuit was subjected to 10 training rounds, with each round consisting of 2000 training iterations.

\begin{figure}[ht]
	\centering
	\includegraphics[width=0.5\textwidth]{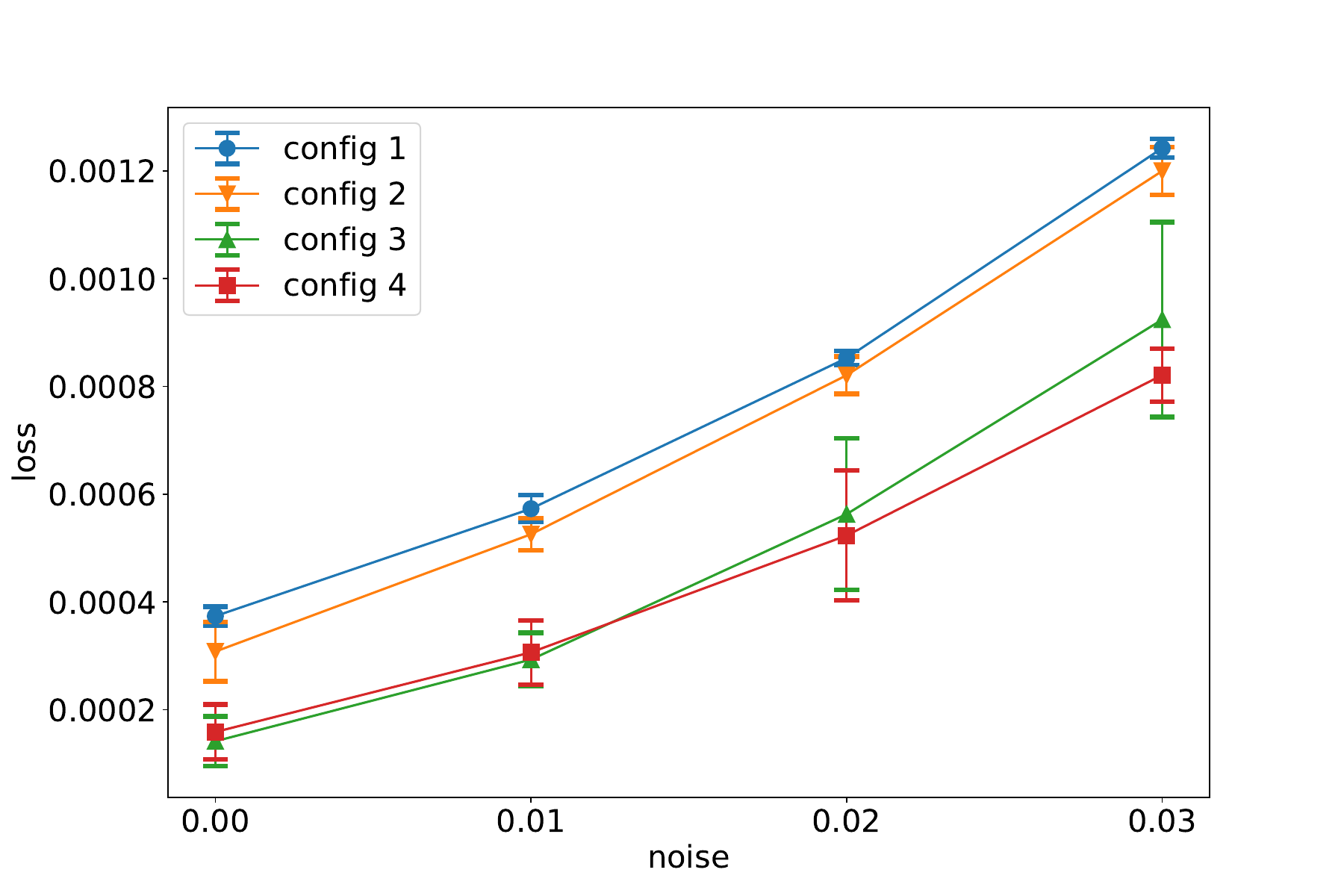}
	\caption{Analysis of VQRA performance under varying noise levels for different circuit configurations.
		This figure presents the loss curves as a function of increasing noise intensity, illustrating how the performance of VQRAs is impacted in noisy environments.
		Each curve represents a different configuration of quantum entanglement within the circuit, providing insights into the resilience of VQRAs against environmental noise
		and the potential role of quantum correlations in mitigating noise effects.}
	\label{fig:noisy_case}
\end{figure}

As depicted in Fig. \ref{fig:noisy_case}, a consistent pattern emerged across all quantum circuit configurations: the loss value escalated as the noise intensity increased. This trend can be attributed to the noise masking the information encoded within the quantum states, thereby diminishing the algorithm's ability to accurately capture the intended data patterns. Interestingly, however, it was observed that configurations allowing for greater quantum correlations demonstrated relatively better resilience to noise. This suggests that the presence of quantum correlations within the circuits contributes to maintaining the VQRA's performance even in noisy conditions, highlighting the robustness of quantum correlations in preserving information integrity.

An interesting observation from our study is that the slope of the loss curve in Fig. \ref{fig:noisy_case} can serve as an indicator of a quantum circuit's resistance to noise. A flatter slope signifies stronger resilience against noise disturbances. In scenarios with low noise intensity ($p < 0.03$), the differences in noise resistance across configurations become evident.
In Configuration 1, where there is an absence of quantum correlations among qubits, the information is localized within individual qubits. This localization makes the information highly susceptible to noise, leading to a sharp increase in the loss curve's slope. This is clearly observed in Fig. \ref{fig:noisy_case} as a rapid incline for Configuration 1.
Contrastingly, Configuration 4, which allows quantum correlations among qubits, demonstrates a more robust response to noise. Here, information is not only stored in individual qubits but also in the correlations between them. During noisy conditions, the VDE learns to encode information into these quantum correlations, providing a buffer against local noise disruptions. This is reflected in Fig. \ref{fig:noisy_case} as a relatively gentler slope for Configuration 4.
However, it's important to note that in high-noise environments, where noise levels are substantial, even non-local quantum correlations are adversely affected, leading to a steep increase in the loss curve's slope. Under such conditions, the VDE struggles to find a secure means of encoding information, and the system's resistance to noise significantly weakens.

\section{Discussion and Conclusion}

In this study, we explored variational data encoding within the framework of VQRAs and examined the pivotal role of quantum correlations in the encoded states. Utilizing PQCs, whose parameters are optimized through machine learning techniques, we successfully demonstrated the efficacy of quantum features in encoding comprehensive information from training data. Our findings suggest that the superposition states in the memory circuit not only store data but may also capture the global structure of the training dataset, offering avenues for quantum enhancement.

Our numerical simulations revealed that VQRAs can achieve remarkable performance even on quantum devices of a limited scale. This showcases the potential of VQRAs in quantum machine learning applications, especially considering the current capabilities of available quantum technology.

Furthermore, we investigated the impact of noise, an inherent characteristic of NISQ devices, on the performance of VQRAs. While noise inevitably impairs the effectiveness of these algorithms, our studies indicate that quantum correlations can significantly bolster the robustness of encoded states against local noise, such as symmetric depolarizing disturbances. This resilience stems from the dual nature of information storage in VQRAs: both in local quantum states and in non-local quantum correlations. Our learning-based encoding schemes adaptively shift towards utilizing quantum correlations in noisy environments, thereby safeguarding the encoded information.

However, our study faced challenges in quantitatively assessing the role of quantum correlations due to the absence of a universally accepted metric for multi-body quantum correlations. Future research endeavors could focus on identifying a suitable measure for quantum correlation and investigating how it varies during training under different noise conditions.

In conclusion, our findings underscore the importance of encoding information into quantum correlations for enhancing quantum machine learning algorithms. Quantum correlations not only provide an expanded space for information storage, potentially conserving quantum resources, but also offer a means to combat local noise. These characteristics are particularly promising for quantum algorithms designed for NISQ devices, suggesting a bright future for quantum-enhanced machine learning.

\vspace{0.3cm}
\section*{Acknowledgments}
\vspace{-0.3cm}

This work was supported by the National Natural Science Foundation of China (under Grant No. 12105090, 12175057).
	
\bibliographystyle{quantum}

%\bibliography{ref2023}

\end{document}